\documentclass[twocolumn,english,superscriptaddress]{revtex4}
\usepackage[T1]{fontenc}
\usepackage[latin9]{inputenc}
\usepackage{amsmath}
\usepackage{amssymb}
\usepackage{stmaryrd}
\usepackage{graphicx}
\usepackage{esint}

\makeatletter
\@ifundefined{textcolor}{}
{%
 \definecolor{BLACK}{gray}{0}
 \definecolor{WHITE}{gray}{1}
 \definecolor{RED}{rgb}{1,0,0}
 \definecolor{GREEN}{rgb}{0,1,0}
 \definecolor{BLUE}{rgb}{0,0,1}
 \definecolor{CYAN}{cmyk}{1,0,0,0}
 \definecolor{MAGENTA}{cmyk}{0,1,0,0}
 \definecolor{YELLOW}{cmyk}{0,0,1,0}
}

\newcommand{\bk}{\mathbf{k}}
\newcommand{\bq}{\mathbf{q}}
\usepackage{babel}

\usepackage{babel}

\usepackage{babel}

\usepackage{babel}

\usepackage{babel}

\usepackage{babel}

\makeatother

\usepackage{babel}
\begin{document}

\title{Distinguishing spin-orbit coupling and nematic order in the electronic
spectrum of iron-based superconductors}

\author{Rafael M. Fernandes}

\affiliation{School of Physics and Astronomy, University of Minnesota, Minneapolis
55455, USA}

\author{Oskar Vafek }

\affiliation{National High Magnetic Field Laboratory and Department of Physics,
Florida State University, Tallahasse, Florida 32306, USA}
\begin{abstract}
The low-energy electronic states of the iron-based superconductors
are strongly affected by both spin-orbit coupling and, when present,
by the nematic order. These two effects have different physical origins,
yet they can lead to similar gap features in the electronic spectrum.
Here we show how to disentangle them experimentally in the iron superconductors
with one Fe plane per unit cell. Although the splitting of the low
energy doublet at the Brillouin zone center ($\Gamma$-point) can
be due to either the spin-orbit coupling or the nematic order, or
both, the degeneracy of each of the doublet states at the zone corner
($M$-point) is protected by the space group symmetry even when spin-orbit
coupling is taken into account. Therefore, any splitting at $M$ must
be due to lowering of the crystal symmetry, such as due to the nematic
order. We further analyze a microscopic tight-binding model with two
different contributions to the nematic order: $d_{xz}/d_{yz}$ onsite
energy anisotropy and the $d_{xy}$ hopping anisotropy. We find that
a precise determination of the former, which has been widely used
to characterize the nematic phase, requires a simultaneous measurement
of the splittings of the $\Gamma$-point doublet and at the two low-energy
$M$-point doublets. We also discuss the impact of twin domains and
show how our results shed new light on ARPES measurements in the normal
state of these materials. 
\end{abstract}
\maketitle

\section{Introduction}

In most iron based superconductors \cite{reviews}, the normal state
displays two instabilities: a spin-density wave (SDW) transition at
$T_{\mathrm{SDW}}$ and an orthorhombic/nematic transition at $T_{\mathrm{nem}}\geq T_{\mathrm{SDW}}$
\cite{Fernandes14}. Angle-resolved photo-emission spectroscopy (ARPES),
being sensitive to the electronic energy-momentum dispersion \cite{Damascelli2003},
is an attractive tool to probe how these distinct ordered states manifest
themselves in the electronic spectrum \cite{OO_Ba122_Shen,OO_NaFeAs_Feng,OO_NaFeAs_Shen,OO_Ba122P_Matsuda,OO_FeSe_Borisenko,OO_FeSe_Shimojima,OO_FeSe_Nakayama,OO_LiFeAs_Ding,OO_FeTeSe_Sugimoto}.
However, the close proximity of the different electronic energy scales,
together with the multi-orbital character of the band structure, render
this task non-trivial \cite{Vafek13}. For instance, in several iron
pnictides, a partial energy gap of about $\mathrm{50}$ meV reported
in optics experiments, and also observed by ARPES at the points where
folded and unfolded bands cross, has been attributed to the formation
of the metallic SDW order \cite{optics_Hu08,optics_Uchida10,Yi14}.
This is of the same order of magnitude as the energy splitting attributed
to the tetragonal symmetry-breaking arising from the formation of
the orthorhombic/nematic phase \cite{OO_Ba122_Shen}. As the system
is doped towards its maximum superconducting transition temperature,
both gaps decrease \cite{Yi14}. Meanwhile, an atomic-like spin-orbit
coupling present in the system gives rise to splittings at the $\Gamma$
point of the order of $10$-$30$ meV in the band structure \cite{SOC_Borisenko14}
without any broken-symmetry \cite{Vafek13}.

Establishing clear criteria to correctly identify the origin of these
spectral features is therefore important to advance our understanding
of the normal state of the iron superconductors. In principle, the
effects of the SDW order -- established independently using neutron
scattering -- on the electronic spectrum can be unambiguously identified,
because, being a commensurate density wave, the lattice translational
symmetry-breaking is manifested in a folding of the band structure
in the momentum space. The case of the nematic splitting is however
more subtle, because it involves only rotational symmetry-breaking,
without the lattice translational symmetry-breaking, and therefore
no zone folding. As an illustration, consider the subspace spanned
by the $d_{xz}$ and $d_{yz}$ Fe orbitals only. On the one hand,
the nematic order breaks the tetragonal symmetry, giving rise to an
additional term in the Hamiltonian: $\Delta_{\mathrm{nem}}\sum_{\mathbf{k}\sigma}\left(c_{xz,\mathbf{k}\sigma}^{\dagger}c_{xz,\mathbf{k}\sigma}^{\phantom{\dagger}}-c_{yz,\mathbf{k}\sigma}^{\dagger}c_{yz,\mathbf{k}\sigma}^{\phantom{\dagger}}\right)$,
where the operator $c_{a,\mathbf{k}\sigma}$ destroys an electron
at orbital $a$ with momentum $\mathbf{k}$ and spin $\sigma$. The
result is the splitting between the on-site energies of these two
orbitals. On the other hand, the spin-orbit coupling $\lambda_{\mathrm{SOC}}$
mixes the two orbitals, and splits the energy of the resulting admixtures,
without breaking the tetragonal symmetry, via the additional term:
$i\lambda_{\mathrm{SOC}}\sum_{\mathbf{k}\sigma}\sigma\left(c_{xz,\mathbf{k}\sigma}^{\dagger}c_{yz,\mathbf{k}\sigma}^{\phantom{\dagger}}-c_{yz,\mathbf{k}\sigma}^{\dagger}c_{xz,\mathbf{k}\sigma}^{\phantom{\dagger}}\right)$.
To distinguish these two features experimentally, one could in principle
use spin polarized ARPES or use the fact that while $\Delta_{\mathrm{nem}}$
has a pronounced temperature dependence, $\lambda_{\mathrm{SOC}}$
is expected to be nearly temperature-independent. However, the electronic
spectral-function's lifetime, manifested as a broadening of the ARPES
data, is also strongly temperature-dependent \cite{Kaminski13,Brouet13},
making this procedure challenging.

It is important to keep in mind that the orbital states are not the
eigenstates of the electronic Bloch problem \cite{Boeri11,w_Ku11,Tomic14,Maier14}.
At first, one may think that this feature renders the qualitative
distinction between the effects of the spin-orbit coupling and the
nematic order less transparent. Indeed, any transformation from the
$3d$ orbital basis of 2 Fe atoms in the crystallographic unit cell
to the band basis requires diagonalizing a $20\times20$ matrix in
the presence of spin-orbit coupling. The size of the matrix is larger
if the puckered pnictogen or chalcogen is taken into account as well.
However, there is an alternative \cite{Vafek13}. Because the Fermi
pockets are usually small and centered at high-symmetry points of
the Brillouin zone, one can focus on the band basis states \textit{directly}
at the high symmetry points. Using the properties of the space-group
representations near these points, we can qualitatively distinguish
the spectral manifestations of spin-orbit coupling and the nematic
splitting, sidestepping the need to analyze large matrices.

\begin{figure}
\begin{centering}
\includegraphics[width=0.9\columnwidth]{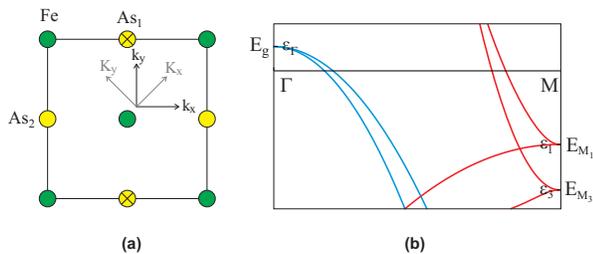} 
\par\end{centering}

\protect\caption{(a) Crystallographic unit cell with 2 Fe atoms. The two coordinate
systems $\left(k_{x},k_{y}\right)$ and $\left(K_{x},K_{y}\right)$
are shown in the figure. (b) Schematic representation of the low-energy
model \cite{Vafek13} with one doublet $E_{g}$ at the $\Gamma=\left(0,0,0\right)$
point and two doublets $E_{M_{1}}$ and $E_{M_{3}}$ at the $M=\left(\pi,\pi,0\right)$
point \cite{doubletnote}. \label{fig_unit_cell}}
\end{figure}

To this end, we use the effective low-energy model derived using the
group-theoretical arguments \cite{Vafek13} in the vicinity of $\Gamma=(0,0,0)$
and $M=(\pi,\pi,0)$ points of the crystallographic Brillouin zone.
Focusing on the iron superconductors with tetragonal $P4/nmm$ symmetry
group, such as the $11$ (FeSe, FeTe), the $111$ (NaFeAs, LiFeAs)
and the $1111$ (LaFeAsO, CeFeAsO) families, and in the presence of
time-reversal symmetry, this model features one doublet at $\Gamma$,
from which two hole pockets originate, and two doublets at $M$, from
which the two electron pockets originate \cite{doubletnote} (see
Fig. \ref{fig_unit_cell}). We find that while the doublet at $\Gamma$
is split by both spin-orbit coupling \cite{Vafek13} and nematic order,
the two doublets at $M$ are split only by the nematic order. Indeed,
without nematic order, the doublets at the $M$ point are guaranteed
\cite{Vafek13}, because each transforms as a single four-dimensional
double-valued irreducible representation of the space group $P4/nmm$
(see Ref. \cite{BradleyCracknell}). This result, being a consequence
of the space group symmetry, is general. Our additional analysis using
the ten-orbital tight-binding model derived from \emph{ab initio}
calculations of Ref. \cite{Eschrig09} shows that one of the doublets
at $M$ is most affected by the on-site energy difference between
the $d_{xz}$ and $d_{yz}$ orbitals, and the other doublet at $M$
by the anisotropy in the hopping parameter connecting the $d_{xy}$
orbitals of neighboring Fe atoms. Because all three quantities contribute
to the splittings of the three doublets, it is necessary to simultaneously
measure the three splittings in ARPES experiments in order to unambiguously
disentangle the effects of nematicity and the spin-orbit coupling.
Finally, we discuss the role played by twin domains and the applicability
of our results to the $122$ family of iron superconductors with $I4/mmm$
space-group symmetry, such as $\mathrm{BaFe_{2}As_{2}}$.

The paper is organized as follows: in Section II we introduce the
low-energy model in the presence of both spin-orbit coupling and nematic
order, and discuss the corresponding band dispersions. In Section
III we compare these results to first-principle calculations fitted
to a $10$-orbital tight-binding model, and to ARPES experiments.
Conclusions are presented in Section IV.

\section{Effective low-energy model}

The low energy effective Hamiltonian for the states near the Fermi
level, corresponding to small hole-like and electron-like pockets,
can be obtained from the $\mathbf{k}.\mathbf{p}$ expansion around
the $\Gamma$ and $M$ points of the crystallographic Brillouin zone,
which contains 2 Fe atoms \cite{Vafek13}. A schematic representation
of the corresponding unit cell is shown in Fig. \ref{fig_unit_cell}a;
hereafter, $xy$ refer to the Fe-As orthogonal directions (parallel
to the crystallographic $ab$ axes), whereas $XY$ refer to the Fe-Fe
orthogonal directions. Here, we focus on the $k_{z}=0$ plane only,
since ARPES measurements have enough resolution to select single $k_{z}$
values via the energy of the incoming photon \cite{OO_FeSe_Borisenko}.
All the orbitals are defined with respect to the Fe-Fe square lattice,
i.e. in the notation $d_{xz}$, $d_{yz}$, $d_{xy}$, $d_{x^{2}-y^{2}}$,
$d_{3z^{2}-r^{2}}$ the subscripts should be understood as referring
to the $XY$ coordinate system.

At $\Gamma$, the irreducible representations of the $P4/nmm$ space
group, suitable for the systems with a single Fe layer per unit cell
(i.e. the $11$, $111$, and $1111$ families) are the same as those
of the well known point group $D_{4h}$. Two hole pockets emerge from
the $E_{g}$ doublet, at an energy $\epsilon_{\Gamma}>0$, which has
degenerate $d_{xz}/d_{yz}$ orbital character (for convenience, we
set the chemical potential to zero). Additional hole pockets may or
may not emerge from the singlets $A_{1g}$ (with $d_{3z^{2}-r^{2}}$
orbital character) and $B_{1g}$ (with $d_{xy}$ character), depending
on their position relative to the Fermi level. Because the main qualitative
changes caused by the nematic order and the spin-orbit coupling occur
in the degenerate states, we focus on the states arising from the
$E_{g}$ doublet only. For each spin projection, $\sigma$, we denote
them by a two component spinor $\psi_{\Gamma,\sigma}\left(\mathbf{k}\right)$.

In contrast, at the $M$ point, the irreducible representations of
the space group cannot be mapped onto those of the point group $D_{4h}$.
This is a consequence of the fact that $P4/nmm$ is a non-symmorphic
space group, due to the presence of an $n$-glide plane symmetry,
i.e. mirror reflection about the Fe plane followed by the translation
by the half unit cell diagonal \cite{Vafek13}. Ignoring the spin
degeneracy, one finds that all irreducible representations at $M$
are two-dimensional, denoted by $E_{M}$ in Ref.\cite{Vafek13}. The
two electron pockets arise from the doublet $E_{M_{1}}$, at an energy
$\epsilon_{1}<0$ and with $d_{xz}/d_{yz}$ orbital character, and
from the doublet $E_{M_{3}}$, at an energy $\epsilon_{3}<\epsilon_{1}<0$
and with $d_{xy}$ orbital character (see Fig. \ref{fig_unit_cell}b).
Out of these 4 states we build two spinors $\psi_{X,\sigma}\left(\mathbf{k}\right)$
and $\psi_{Y,\sigma}\left(\mathbf{k}\right)$, whose upper (lower)
components transform respectively as $E_{M_{1}}^{X}$ and $E_{M_{1}}^{Y}$
($E_{M_{3}}^{X}$ and $E_{M_{3}}^{Y}$). A schematic representation
of these states is shown in Fig. \ref{fig_unit_cell}.

Defining the enlarged spinor $\Psi_{\sigma}^{\dagger}=\left(\begin{array}{ccc}
\psi_{X,\sigma}^{\dagger} & \psi_{Y,\sigma}^{\dagger} & \psi_{\Gamma,\sigma}^{\dagger}\end{array}\right)$, the low-energy Hamiltonian can be written as 
\begin{equation}
\mathcal{H}=\sum_{\mathbf{k}\alpha\beta}\Psi_{\alpha}^{\dagger}\left(\mathbf{k}\right)H_{\alpha\beta}\left(\mathbf{k}\right)\Psi_{\beta}^{\phantom{\dagger}}\left(\mathbf{k}\right),
\end{equation}
where (suppressing $\alpha,\beta$ and $\bk$) the matrix 
\begin{equation}
H=H_{0}+H_{\mathrm{SOC}}+H_{\mathrm{nem}},
\end{equation}
is the single-particle Hamiltonian in the nematic-paramagnetic state.
The first two terms were obtained in Ref. \cite{Vafek13} and we repeat
here the results. For the non-interacting $H_{0}$ part, we obtain:
\begin{equation}
H_{0}\left(\mathbf{k}\right)=\left(\begin{array}{ccc}
h_{M}^{+}\left(\mathbf{k}\right) & 0 & 0\\
0 & h_{M}^{-}\left(\mathbf{k}\right) & 0\\
0 & 0 & h_{\Gamma}\left(\mathbf{k}\right)
\end{array}\right)\varotimes\sigma_{0},\label{H0}
\end{equation}
with $2\times2$ matrices: 
\begin{align}
h_{M}^{\pm}\left(\mathbf{k}\right) & =\sum_{i=1,3}\left(\epsilon_{i}+\frac{k^{2}}{2m_{i}}\pm a_{i}k_{x}k_{y}\right)\tilde{\tau}_{i}+v_{\pm}\left(\mathbf{k}\right)\tau_{2},\nonumber \\
h_{\Gamma}\left(\mathbf{k}\right) & =\left(\epsilon_{\Gamma}+\frac{k^{2}}{2m_{\Gamma}}\right)\tau_{0}+bk_{x}k_{y}\tau_{3}+c\left(k_{x}^{2}-k_{y}^{2}\right)\tau_{1},\label{aux_H0}
\end{align}
where $\tilde{\tau}_{1}=\frac{1}{2}\left(\tau_{0}+\tau_{3}\right)$,
$\tilde{\tau}_{3}=\frac{1}{2}\left(\tau_{0}-\tau_{3}\right)$, and:

\begin{align}
v_{\pm}\left(\mathbf{k}\right) & =v\left(\pm k_{x}+k_{y}\right)+p_{1}\left(\pm k_{x}^{3}+k_{y}^{3}\right)\nonumber \\
 & +p_{2}k_{x}k_{y}\left(k_{x}\pm k_{y}\right)\label{aux_v}
\end{align}

\begin{figure}
\begin{centering}
\includegraphics[width=0.9\columnwidth]{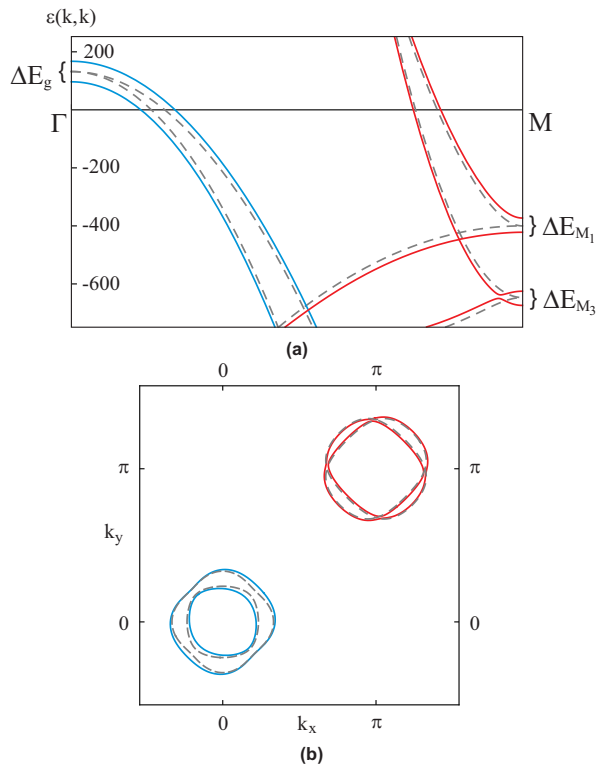} 
\par\end{centering}

\protect\caption{(a) Band dispersion along the $\Gamma$-$M$ line in units of meV
and (b) Fermi surface of the effective low-energy model in the presence
(solid lines) and in the absence (dashed lines) of both spin-orbit
coupling and nematic order (solid lines). Red (blue) lines denote
the states emerging from the $M$ ($\Gamma$) doublets, which will
give rise to electron (hole) pockets. The doublet splittings are explicitly
shown. In this figure, the parameters used were $\lambda=\varphi_{i}=50$
meV. \label{fig_dispersions}}
\end{figure}

The Pauli matrices $\sigma$ refer to the spin space, whereas the
Pauli matrices $\tau$ refer to the spinor space. The simplicity and
generality of this model should be evident when compared to the $10$-orbital
tight-binding model with fifth-neighbor hopping parameters. The $13$
free parameters are material-dependent, and can be fit to first-principle
calculations. For concreteness, in this paper we use the parameters
defined in the first row of Table IX of Ref. \cite{Vafek13}. The
corresponding band dispersions and Fermi surfaces are shown as dashed
lines in Fig. \ref{fig_dispersions}. For the spin-orbit term $H_{\mathrm{SOC}}$
we have:

\begin{equation}
H_{\mathrm{SOC}}\left(\mathbf{k}\right)=\left(\begin{array}{ccc}
0 & h_{M}^{\mathrm{SOC}}\left(\mathbf{k}\right) & 0\\
\left(h_{M}^{\mathrm{SOC}}\right)^{\dagger}\left(\mathbf{k}\right) & 0 & 0\\
0 & 0 & h_{\Gamma}^{\mathrm{SOC}}\left(\mathbf{k}\right)
\end{array}\right),\label{H_SOC}
\end{equation}
with $4\times4$ matrices: 
\begin{align}
h_{M}^{\mathrm{SOC}}\left(\mathbf{k}\right) & =i\frac{\lambda}{4}\left(\tau_{+}\varotimes\sigma_{1}+\tau_{-}\varotimes\sigma_{2}\right),\nonumber \\
h_{\Gamma}^{\mathrm{SOC}}\left(\mathbf{k}\right) & =\frac{\lambda}{2}\,\tau_{2}\varotimes\sigma_{3},\label{aux_HSOC}
\end{align}
and using the usual definition $\tau_{\pm}=\tau_{1}\pm i\tau_{2}$.
Here, $\lambda$ sets the strength of the spin-orbit coupling, which,
without loss of generality, we take to be the same near $\Gamma$
and $M$.

To derive the nematic term $H_{\mathrm{nem}}$, which breaks the tetragonal
symmetry along the Fe-Fe directions, it is enough to find combinations
of components of $\Psi_{\sigma}(\bk)$ and $\bk$, which transform
as $B_{2g}$ invariants, i.e. as $k_{x}k_{y}\propto K_{x}^{2}-K_{y}^{2}$.
Clearly, there can be no bilinear terms which mix the two-component
spinors at $\Gamma$ and $M$, because such terms would break the
lattice translational symmetry. Meanwhile, nematic is a $\bq=0$ order.
The $\bk$ independent terms near $\Gamma$ follow immediately from
the form of $h_{\Gamma}(\bk)$: the only combination of the two components
of $\psi_{\Gamma,\sigma}$ which transform as $k_{x}k_{y}$ is the
combination which multiplies this term in $h_{\Gamma}(\bk)$, because
the resulting combination is an invariant, i.e. transforms trivially.
Therefore, near $\Gamma$, the nematic order induces a $\bk$-independent
term $\sim\sum_{\bk,\sigma}\psi_{\Gamma,\sigma}^{\dagger}(\bk)\tau_{3}\psi_{\Gamma,\sigma}(\bk)$.
Similarly, at $M$, the $\bk$-independent combinations which transform
as $B_{2g}$ can be read off from $h_{M}^{\pm}(\bk)$. The two independent
terms are $\sim\sum_{\bk,\sigma}\left(\psi_{X,\sigma}^{\dagger}(\bk)\tilde{\tau}_{i}\psi_{X,\sigma}(\bk)-\psi_{Y,\sigma}^{\dagger}(\bk)\tilde{\tau}_{i}\psi_{Y,\sigma}(\bk)\right)$,
where $i=1$ or $i=3$. We then obtain 
\begin{equation}
H_{\mathrm{nem}}\left(\mathbf{k}\right)=\left(\begin{array}{ccc}
h_{M}^{\mathrm{nem}}\left(\mathbf{k}\right) & 0 & 0\\
0 & -h_{M}^{\mathrm{nem}}\left(\mathbf{k}\right) & 0\\
0 & 0 & h_{\Gamma}^{\mathrm{nem}}\left(\mathbf{k}\right)
\end{array}\right)\varotimes\sigma_{0},\label{Hnem}
\end{equation}
with $2\times2$ matrices: 
\begin{align}
h_{M}^{\mathrm{nem}}\left(\mathbf{k}\right) & =\frac{\varphi_{1}}{4}\left(\tau_{0}+\tau_{3}\right)+\frac{\varphi_{3}}{4}\left(\tau_{0}-\tau_{3}\right),\nonumber \\
h_{\Gamma}^{\mathrm{nem}}\left(\mathbf{k}\right) & =\frac{\varphi_{\Gamma}}{2}\,\tau_{3}.\label{aux_H_nem}
\end{align}

Symmetry alone is unable to fix the values of $\varphi_{j}$. Nevertheless,
because all $\varphi_{j}$'s are nematic order parameters, they must
be related to each other microscopically, although they do not need
to be \textit{equal} to each other. It is now straightforward to obtain
the electronic spectrum in the presence of both nematic order and
spin-orbit coupling. We find splittings in the $E_{g}$ doublet at
the $\Gamma$ point (energy $\epsilon_{\Gamma}>0$), as well as in
the $E_{M_{1}}$ and $E_{M_{3}}$ doublets at the $M$ point (energies
$0<\epsilon_{1}<\epsilon_{3}$), given by:

\begin{align}
\Delta E_{g} & =\sqrt{\lambda^{2}+\varphi_{\Gamma}^{2}}\nonumber \\
\Delta E_{M_{1/3}} & =\pm\left(\frac{\varphi_{1}-\varphi_{3}}{2}\right)-\frac{1}{2}\sqrt{\lambda^{2}+\left(\epsilon_{1}-\epsilon_{3}-\frac{\varphi_{1}+\varphi_{3}}{2}\right)^{2}}\nonumber \\
 & +\frac{1}{2}\sqrt{\lambda^{2}+\left(\epsilon_{1}-\epsilon_{3}+\frac{\varphi_{1}+\varphi_{3}}{2}\right)^{2}}.\label{aux_splittings}
\end{align}
To make the expressions of the splittings at the $M$ point more transparent,
it is useful to consider an expansion in powers of $\varphi$: 
\begin{align}
\Delta E_{g} & =\sqrt{\lambda^{2}+\varphi_{\Gamma}^{2}},\nonumber \\
\Delta E_{M_{1}} & =\varphi_{1}-\frac{\left(\varphi_{1}+\varphi_{3}\right)\lambda^{2}}{4\left(\epsilon_{1}-\epsilon_{3}\right)^{2}}+\mathcal{O}\left(\lambda^{2}\varphi^{3}\right),\nonumber \\
\Delta E_{M_{3}} & =\varphi_{3}-\frac{\left(\varphi_{1}+\varphi_{3}\right)\lambda^{2}}{4\left(\epsilon_{1}-\epsilon_{3}\right)^{2}}+\mathcal{O}\left(\lambda^{2}\varphi^{3}\right).\label{splittings}
\end{align}
These splittings, the corresponding band dispersions, and the Fermi
surface distortions are shown as solid lines in Fig. \ref{fig_dispersions}.
The fact that the splitting of the high-symmetry doublets are not
simply proportional to the nematic order parameter, but depend also
on the spin-orbit coupling, is one of our main results. Eq. (\ref{splittings})
shows that the $E_{g}$ doublet at the $\Gamma$ point can be split
by spin-orbit coupling even without tetragonal symmetry-breaking.
Moreover, the splitting is insensitive to the sign of the nematic
order parameter $\varphi_{\Gamma}$. We also note that the $E_{M}$
doublets at the $M$ point can only be split by nematic order. However,
once they are split, the spin-orbit coupling also contributes to the
splitting, and the magnitude of this contribution depends also on
the proximity between the unperturbed doublets, $\left|\epsilon_{1}-\epsilon_{3}\right|$.
Furthermore, while $\varphi_{1}\neq0$ gives the dominant splitting
of $E_{M_{1}}$, $\varphi_{3}$ can also cause a splitting in $E_{M_{1}}$
even if $\varphi_{1}=0$, as long as the spin-orbit coupling is non-zero.

Interestingly, the most prominent manifestations of nematic order
and spin-orbit coupling on the Fermi surface occur in different pockets
(Fig. \ref{fig_dispersions}). On the one hand, nematic order has
a stronger effect on the hole pockets, as $\varphi_{\Gamma}>0$ ($\varphi_{\Gamma}<0$)
causes the two hole pockets to ``repel'' each other predominantly
along the $K_{x}$ ($K_{y}$) direction. On the other hand, spin-orbit
coupling avoids the crossing between the two electron pockets, giving
rise to two separate pockets. We emphasize again, despite the presence
of spin-orbit coupling and the nematic order parameter, all bands
remain doubly-degenerate due to the presence of time reversal and
a center of space inversion \cite{Vafek13,doubletnote}.

\section{Comparison to first-principle calculations and ARPES experiments}

The results derived above are general and independent of the particularities
of the band structure. They rely only on the space group symmetry
$P4/nmm$ and its consequences at the doublets at $\Gamma$ and $M$.
Yet, in order to understand the microscopic meaning of the nematic
order parameters $\left(\varphi_{\Gamma},\varphi_{1},\varphi_{3}\right)$,
it is instructive to compare our results with those obtained from
10-orbital tight-binding models fitted to first-principle calculations.
We used the model of Ref. \cite{Eschrig09} for LaFeAsO \cite{TBnote}
and added two different tetragonal symmetry-breaking terms, while
ignoring the spin-orbit coupling. The first is a uniform onsite energy
splitting between the $d_{xz}$ and $d_{yz}$ orbitals of the form:
\begin{equation}
H_{1}=\frac{\Delta_{1}}{2}\sum_{\mathbf{k}\sigma}\left(c_{xz,\mathbf{k}\sigma}^{\dagger}c_{xz,\mathbf{k}\sigma}^{\phantom{\dagger}}-c_{yz,\mathbf{k}\sigma}^{\dagger}c_{yz,\mathbf{k}\sigma}^{\phantom{\dagger}}\right),\label{aux_H1}
\end{equation}
whereas the second is a hopping anisotropy between the $d_{xy}$ orbitals
of the two Fe atoms living in the same unit cell: 
\begin{equation}
H_{3}=\frac{\Delta_{3}}{2}\sum_{\mathbf{k}\sigma}c_{xy(1),\mathbf{\mathbf{k}}\sigma}^{\dagger}c_{xy(2),\mathbf{\mathbf{k}}\sigma}^{\phantom{\dagger}}\sin\left(\frac{k_{x}}{2}\right)\sin\left(\frac{k_{y}}{2}\right).\label{aux_H3}
\end{equation}

\begin{figure}
\begin{centering}
\includegraphics[width=0.8\columnwidth]{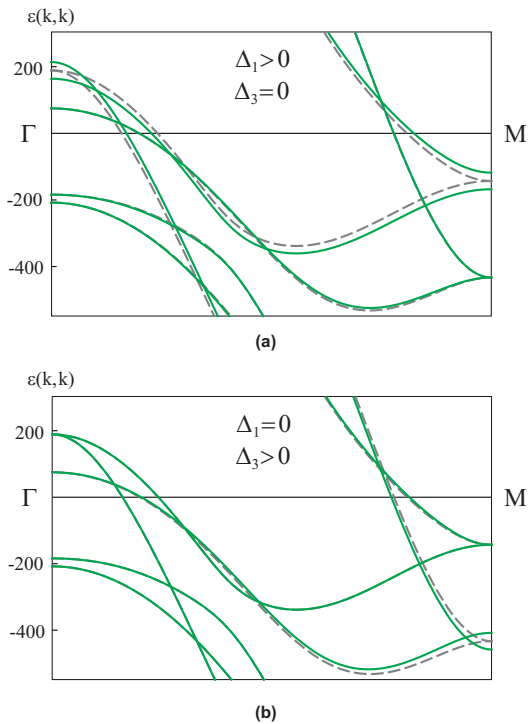} 
\par\end{centering}

\protect\caption{Band dispersion along the $\Gamma$-$M$ direction (in meV) for the
10-orbital model of Ref. \cite{Eschrig09} in the presence of (a)
an onsite energy anisotropy between the $d_{xz}/d_{yz}$ orbitals,
$\Delta_{1}=50$ meV, and (b) a hopping anisotropy between $d_{xy}$
orbitals of nearest-neighbor Fe atoms, $\Delta_{3}=50$ meV. Dashed
lines represent the dispersions for $\Delta_{1}=\Delta_{3}=0$. \label{fig_TB_model}}
\end{figure}

The results shown in Fig. \ref{fig_TB_model} for $\Delta_{1}>0$,
$\Delta_{3}=0$ and $\Delta_{1}=0$, $\Delta_{3}>0$ reveal that whereas
the $E_{g}$ and the $E_{M_{1}}$ doublets are split by $\Delta_{1}$,
the $E_{M_{3}}$ doublet is split by $\Delta_{3}$, with $\Delta E_{g}=\Delta E_{M_{1}}=\Delta_{1}$
and $\Delta E_{M_{3}}=\Delta_{3}$. This is not unexpected, since
in our low-energy model the $E_{g}$ and $E_{M_{1}}$ doublets have
$d_{xz}/d_{yz}$ characters, whereas the $E_{M_{3}}$ doublet has
$d_{xy}$ character \cite{Vafek13}. Therefore, we can identify $\varphi_{\Gamma}=\varphi_{1}$
with the on-site $d_{xz}/d_{yz}$ orbital anisotropy $\Delta_{1}$
and $\varphi_{3}$ to the $d_{xy}$ hopping anisotropy $\Delta_{3}$.
Note that these results are insensitive to the microscopic mechanism
behind the tetragonal symmetry breaking, i.e. whether it arises due
to orbital \cite{kruger09,w_ku10,Devereaux10,Phillips11,Phillips12,Kontani12,Nevidomskyy14,Littlewood}
or spin fluctuations \cite{Fernandes14,Kivelson08,Sachdev08,Fernandes12,Dagotto13,Kang14},
or electron-phonon coupling.

\begin{figure}
\begin{centering}
\includegraphics[width=0.9\columnwidth]{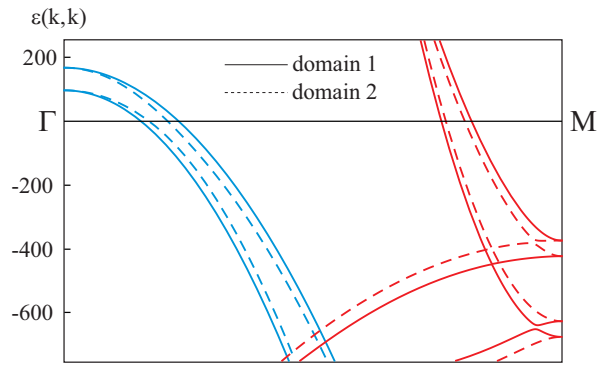} 
\par\end{centering}

\protect\caption{Band dispersions along the $\Gamma$-$M$ direction (in meV) for two
different types of nematic domain. Solid (dashed) lines correspond
to the domain with $\varphi_{i}=50$ meV ($\varphi_{i}=-50$ meV).
In both cases, $\lambda=50$ meV. \label{fig_domains}}
\end{figure}

In making quantitative comparison with experiments, an important issue
is the presence of twin domains, i.e. while certain regions of the
sample are characterized by a set of nematic order parameters $\left(\varphi_{\Gamma},\varphi_{1},\varphi_{3}\right)$,
other regions display $\left(-\varphi_{\Gamma},-\varphi_{1},-\varphi_{3}\right)$.
To take this effect into account, in Fig. \ref{fig_domains} we superimpose
the band dispersions of the two different types of domains using our
effective model of the previous Section. Since ARPES averages over
the entire sample, in principle it should be capable of observing
the two sets of band dispersions as long as the momentum and energy
resolutions are large enough. As expected from symmetry considerations,
and confirmed by Eq. (\ref{splittings}), the magnitudes of the splittings
are insensitive to the sign of the nematic order parameters. In contrast,
the band dispersions away from the high-symmetry points are different
for distinct domains, resulting in an effective doubling of the number
of bands -- even though the translational symmetry is not lowered.
Remarkably, the hole-pockets dispersions are very similar for the
two domain types, reflecting the dependence of the $E_{g}$ splitting
on $\varphi_{\Gamma}^{2}$. Thus, depending on the ARPES resolution,
it may be difficult to resolve the different domain contributions
near the $\Gamma$ point. Meanwhile, the electron-pockets dispersions
are strongly dependent on the sign of $\varphi_{1}$ and $\varphi_{3}$.
In particular, for one sign of the nematic order parameter, the two
branches dispersing from a given $E_{M_{i}}$ doublet either ``attract''
or ``repel'' each other -- although spin-orbit coupling in general
leads to avoided level crossing. As a result, it is more likely for
ARPES to be able to resolve the two domains contributions near the
$M$ point.

We can now discuss the implications of our results to the interpretation
of ARPES experiments in the nematic-paramagnetic state, which takes
place in the temperature range $T_{\mathrm{SDW}}<T<T_{\mathrm{nem}}$
for unstrained samples and in the temperature range $T_{\mathrm{SDW}}<T$
for detwinned samples \cite{OO_Ba122_Shen,OO_NaFeAs_Feng,OO_NaFeAs_Shen,OO_Ba122P_Matsuda,OO_FeSe_Borisenko,OO_FeSe_Shimojima,OO_FeSe_Nakayama,OO_LiFeAs_Ding,OO_FeTeSe_Sugimoto}.
The observed splitting of the $E_{M_{1}}$ doublet has been mostly
attributed to an anisotropy in the onsite energies of the $d_{xz}$
and $d_{yz}$ orbitals (ferro-orbital order), which in our model is
given by $\varphi_{\Gamma}=\varphi_{1}$. However, our results in
Eq. (\ref{splittings}) show that this splitting depends also on the
nematic order parameter $\varphi_{3}$, corresponding to anisotropic
$d_{xy}$ hopping, and on the spin-orbit coupling $\lambda$. Thus,
to properly disentangle these three contributions $\varphi_{1}=\varphi_{\Gamma}$,
$\varphi_{3}$, and $\lambda$ we argue that it is necessary to measure
simultaneously the splitting of the two doublets at the $M$ point
and of the doublet at the $\Gamma$ point. The interplay between these
three parameters may also explain why the doublet splittings are different
at these two high-symmetry points, as observed experimentally \cite{Yi14}.

Applied to twin samples, our results reveal important distinctions
in the twin-domains dispersions near the $\Gamma$ and $M$ points,
as shown in Fig. \ref{fig_domains}. Specifically, while at $\Gamma$
the dispersions are very similar for the two twin domains, at $M$
they are significantly different, since only in the latter the nematic
order gives rise to a bonding and an anti-bonding orbital mixing.
Moreover, the doubling of the number of bands is not a consequence
of translational symmetry-breaking, since the nematic order is a $\mathbf{q}=0$
order.

Finally, we comment on the application of our results to the 122 materials,
whose space group is $I4/mmm$ instead of $P4/nmm$. In contrast to
the latter, the former space group is symmorphic\cite{Vafek13}. In
the limit of no coupling between the Fe layers, this distinction is
irrelevant, and the results derived here for the 1111, 111, and 11
families would also apply to the 122 family. Turning on a weak inter-layer
coupling should lead to small changes, mixing states at $k_{z}=\pi$
to $k_{z}=0$. Although such an additional mixing could complicate
the analysis proposed here, it remains to be seen whether it can be
resolved by current ARPES experiments.

\section{Conclusions}

In summary, we used a low-energy model that respects all symmetries
of the $P4/nmm$ iron supercondutors to reveal the interplay between
nematic order and spin-orbit coupling in the electronic spectrum of
these materials. The simple expressions obtained for the splittings
of the three doublets located at the high-symmetry points of the crystallographic
Brillouin zone, Eq. (\ref{splittings}), enables one to distinguish
unambiguously using ARPES experiments not only these two physical
effects, but also the two different contributions to the nematic order
-- namely the $d_{xz}/d_{yz}$ orbital and $d_{xy}$ hopping anisotropies.
These criteria to disentangle spin-orbit and nematicity, being independent
of details of the band dispersions, open an interesting route to systematically
study the energy scales and the relevance of these two physical effects
to the normal state of different iron-based superconductor families.

We thank S. Borisenko, A. Chubukov, V. Cvetkovic, I. Eremin, and R.
Valenti for fruitful discussions. RMF is supported by the Department
of Energy under Award Number DE -- SC0012336. OV is supported by the
NSF CAREER award under Grant No. DMR-0955561. We would also like to
thank the KITP-UCSB Research Program, ``Magnetism, Bad Metals and
Superconductivity'', where this work was initiated, for hospitality.
KITP is supported in part by NSF Grant No. NSF PHY11-25915.


\begin{thebibliography}{10}
\bibitem{doubletnote} We should mention, that without spin-orbit,
each ``doublet'' has an additional two-fold spin degeneracy. This
remnant degeneracy survives inclusion of either spin-orbit coupling
or nematic, or both; such two-fold degeneracy is protected by the
presence of time reversal symmetry and a center of space inversion\cite{Vafek13}.
Despite this additional degeneracy, we will continue referring to
them as ``doublets''.
\bibitem{TBnote} To remove an artificial degeneracy of the 10-orbital
tight-binding model of Ref. \cite{Eschrig09}, and be consistent with
the first-principle results, we added a small (symmetry allowed) nearest-neighbor
hopping between the two $z^{2}$ orbitals in the unit cell, $2\tilde{t}\sum_{\mathbf{k}\sigma}c_{z^{2}(1),\mathbf{\mathbf{k}}\sigma}^{\dagger}c_{z^{2}(2),\mathbf{\mathbf{k}}\sigma}^{\phantom{\dagger}}\left[\cos\left(\frac{k_{x}+k_{y}}{2}\right)+\cos\left(\frac{k_{x}-k_{y}}{2}\right)\right]$
with $\tilde{t}=-3$ meV.
\end{thebibliography}
\end{document}